# The cluster Terzan 5 as a remnant of a primordial building block of the Galactic bulge

F.R. Ferraro[1], E. Dalessandro[1], A. Mucciarelli[1], G.Beccari[2], M. R. Rich[3], L. Origlia[4], B. Lanzoni[1], R. T. Rood[5], E. Valenti[6,7], M. Bellazzini[4], S. M. Ransom[8], G. Cocozza[4]

*[1]Department of Astronomy, University of Bologna, Via Ranzani, 1, 40127 Bologna, Italy*

*[2]ESA, Space Science Department, Keplerlaan 1, 2200 AG Noordwijk, Netherlands*

*[3]Department of Physics and Astronomy, Math-Sciences 8979, UCLA, Los Angeles, CA 90095-1562,USA*

*[4]INAF- Osservatorio Astronomico di Bologna, Via Ranzani, 1, 40127 Bologna, Italy*

*[5]Astronomy Department, University of Virginia, P.O. Box 400325, Charlottesville, VA, 22904,USA*

*[6]European Southern Observatory, Alonso de Cordova 3107, Vitacura, Santiago, Chile*

*[7]Pontificia Universidad Catolica de Chile, Departamento de Astronomia, Avda Vicuña Mackenna 4860, 782-0436 Macul, Santiago, Chile*

*[8]National Radio Astronomy Observatory, Charlottesville, VA 22903, USA*

**Globular star clusters are compact and massive stellar systems old enough to have witnessed the entire history of our Galaxy, the Milky Way. Although recent results[1,2,3] suggest that their formation may have been more complex than previously thought, they still are the best approximation to a stellar population formed over a relatively short time scale (less than 1 Gyr) and with virtually no dispersion in the iron content. Indeed, only one cluster-like system (ω Centauri) in**

**the Galactic halo is known to have multiple stellar populations with a significant spread in iron abundance and age[4,5]. Similar findings in the Galactic bulge have been hampered by the obscuration arising from thick and varying layers of interstellar dust. Here we report that Terzan 5, a globular-cluster-like system in the Galactic bulge, has two stellar populations with different iron content and ages. Terzan 5 could be the surviving remnant of one of the primordial building blocks that are thought to merge and form galaxy bulges.**

We have recently obtained a set of high-resolution images of Terzan 5 in the *K* and *J* bands by using MAD[6], a Multi-Conjugate Adaptive Optics demonstrator instrument installed at the Very Large Telescope (VLT) of the European Southern Observatory (ESO). MAD operates at near-infrared wavelengths, thus revealing the only component of stellar radiation that can efficiently cross the thick clouds of dust obscuring the Galactic bulge. It is able to perform exceptionally good and uniform adaptive optics correction over its entire field of view (1'x1'), thus compensating for the degradation effects to the astronomical images induced by the Earth's atmosphere. In particular, we have obtained a set of *K*-band (2.2μm) images of Terzan 5 close to the diffraction limit (Fig. 1). The sharpness and uniformity of the images yields very high quality photometry, resulting in accurate (*K, J–K*) colour-magnitude diagram (CMD) even for the very central region of the cluster, and leading to a surprising discovery. We have detected two well-defined red horizontal branch clumps, separated in luminosity: a bright horizontal branch (*BHB*) at $K = 12.85$ and a faint horizontal branch (*FHB*) at $K = 13.15$, the latter having a bluer *(J–K)* colour (Fig. 2).

We have carefully considered whether the double horizontal branch could be spurious. It is neither due to instrumental effects (Fig. 2), nor to differential reddening[7,8] (as the two horizontal branch clumps in the CMD are separated in a direction which is essentially orthogonal to the reddening vector), nor to field contamination (while field stars are expected to be almost uniformly distributed over the MAD field of view, the radial distributions of the stars belonging to the two horizontal branch clumps are significantly concentrated toward the cluster centre and are inconsistent with a uniform distribution at more than $5\sigma$ level; see Fig. 3a and Supplementary Information). We have also found that the radial distributions of the two horizontal branch populations are different (Fig. 3a): according to a Kolmogorov-Smirnov test, the *BHB* population is significantly (at $> 3.5\sigma$ level) more centrally concentrated than that of the *FHB*. The stars belonging to the *BHB* are also substantially more numerous than those of the *FHB* near the cluster centre (that is, at distances $r < 20''$), becoming progressively more rare at larger radii (Fig. 3b).

Once alerted to the existence of the double horizontal branch, we have also identified the feature in optical observations obtained with the Advanced Camera for Surveys (ACS) on board the Hubble Space Telescope (HST; see Supplementary Fig. 1a). Although the strong differential reddening broadens the colour extension of the horizontal branch clumps by ~ 1 mag, the optical ($I$, $V-I$) CMD still shows a clear bimodal distribution of horizontal branch stars in the direction orthogonal to the reddening vector (Supplementary Fig. 1b). A hint of a double horizontal branch clump was already visible in a previously published CMD obtained with HST-NICMOS[9-11], although the shorter colour baseline provided by the *J*- and *H*- band observations did not clearly separate the two clumps.

Hence, we conclude that the existence of the two horizontal branch clumps is a real feature, and the differing radial distributions may indicate different physical origins of the two populations. In particular, a combination of different metallicity and age, with the population in the *BHB* clump being more metal-rich and younger than that in the *FHB* clump, could in principle reproduce the observed features (Supplementary Fig. 2). The only direct information previously available on the metal content of individual stars in Terzan 5 was from four bright giants near the Tip of the red giant branch (RGB), giving an average iron-to-hydrogen abundance ratio [Fe/H] = −0.2 with a negligible dispersion[12]. Hence, we quickly secured medium-resolution near-infrared spectra of 6 horizontal branch stars (3 in each clump) at the Keck Telescope[13]. The derived radial velocities for the two groups of stars (−85 km $s^{-1}$ in both cases) are fully consistent with the previous measures[12] and the systemic velocity of Terzan 5 quoted in the currently adopted globular cluster catalogue[14]. This confirms that all of the observed stars are cluster members and suggests that there is no significant kinematical difference between the two populations (this is also confirmed by proper motion studies; see Supplementary Information). Furthermore, we have found that the iron content of the stars in the two clumps differs by a factor of 3 (~ 0.5 dex): the *FHB* stars have [Fe/H] = − 0.2, while the *BHB* stars have [Fe/H] = +0.3 (Fig. 4a).

To date, apart from a significant spread in the abundance patterns of a few light elements (such as Na and O)[1], the chemical composition of all globular clusters in the Galaxy is known to be extremely uniform in terms of iron content, with the only exception being ω Centauri[4,5] in the Galactic halo. Hence, Terzan 5 is the first stellar aggregate discovered in the Galactic bulge that has globular-cluster-like properties but also with the signatures of a much more complex star formation history.

To further investigate this issue, we have performed a differential reddening correction[15] on the optical ACS catalogue and combined it with the near-infrared data, thus obtaining the (*K, V*–*K*) CMD shown in Fig. 4b. The presence of two distinct populations with a double horizontal branch and (possibly) two separate RGBs can be seen in this CMD. The RGB of the most metal-rich population appears to be more bent (as expected, because of the line blanketing due to a higher metal content). The observed features can be reproduced with two populations characterized by the observed metallicities and two different ages: $t$ = 12 Gyr for the *FHB* and a significantly younger age ($t$ = 6 Gyr) for the *BHB*.

Using the number of horizontal branch stars found in the combined MAD and ACS samples (see Supplementary Information for details), we estimate that the cluster harbours about 800 *FHB* stars and 500 *BHB* stars in total. This is even larger than the global horizontal branch population of 47 Tucanae[16], thus suggesting that Terzan 5 is more massive than previously thought (Supplementary Information).

The evidence for two distinct stellar populations and for a very large total mass suggests that Terzan 5 has experienced a quite troubled formation history. It might be the merger-product of two independent stellar aggregates[17]. Although such a possibility seems to be unlikely for globular clusters belonging to the Galactic halo, the chance of capturing a completely independent stellar system should be significantly larger if the orbits are confined within the Galactic bulge. In this scenario, however, it is not easy to explain why the metal-rich population is more centrally concentrated than the metal-poor one. Moreover, globular clusters younger than 10 Gyr are very rare in our Galaxy[18]. Rather, Terzan 5 could be a complex ω Centauri-like system[4,5] or the nuclear remnant of a disrupted galaxy, similar to the M 54-Sagittarius system[19,20] or the Carina dwarf spheroidal[21] in the metal-rich regime. The remnant of a disrupted dwarf galaxy would naturally present a larger central concentration of the metal-rich (and younger) population[22], as commonly observed in the satellites of the Milky Way and M31. On the other hand, the strict similarity in iron abundance between Terzan 5 and the Galactic bulge population is fully compatible with the hypothesis that the (partial) disruption of its progenitor has contributed to the formation of the Galactic bulge[23].

Possible relics of the hierarchical assembly of the Galactic halo have been recently identified at high Galactic latitudes[24]. Terzan 5 may be the first example of the sub-structures that contributed to form the Galactic bulge. Indeed, our discovery could be the observational confirmation that galactic spheroids originate from the merging of pre-formed, internally evolved stellar systems, and that other similar objects might be hidden in the heavily obscured central region of the Galaxy.

**Supplementary Information** accompanies the paper on **www.nature.com/nature**.

**Acknowledgements:** We thank the MAD team at ESO and in particular P. D'Amico for performing the observations with MAD. This research was supported by "Progetti Strategici di Ateneo 2006 " ( University of Bologna), "Progetti di Ricerca di Interesse Nazionale 2007 and 2008" (Istituto Nazionale di Astrofisica), Agenzia Spaziale Italiana and the Ministero dell'Istruzione, dell'Universitá e della Ricerca. We also acknowledge support from the ESTEC Faculty Visiting Scientist Programme. R.M.R. is supported by the NSF and STScI; R.T.R. is partially supported by STScI. This research has made use of the ESO/ST-ECF Science Archive facility, which is a joint collaboration of the ESO and the Space Telescope - European Coordinating Facility. Part of the data presented here were obtained at the W.M. Keck Observatory, which is operated as a scientific partnership among the California Institute of Technology, the University of California and the NASA. The Observatory was made possible by the financial support of the W.M. Keck Foundation.

**Author contributions:** F.R.F. designed the study and coordinated the activity. E.D., A.M. G.B., E.V. and G.C. analysed the photometric dataset. R.M.R. and L.O. secured and analyzed the Keck spectra. A.M. designed the reddening correction routine. M.B. performed radial distributions tests. E.D. and M.B performed the proper motion analysis. F.R.F., B.L. and L.O. wrote the paper. S.R., R.M.R., R.T.R. and M.B. critically contributed to the paper presentation. All the authors contributed to discussion of the results and commented on the manuscript.

**Correspondence to: F.R.Ferraro[1] Correspondence and requests for materials should be addressed to F.R.F. francesco.ferraro3@unibo.it**

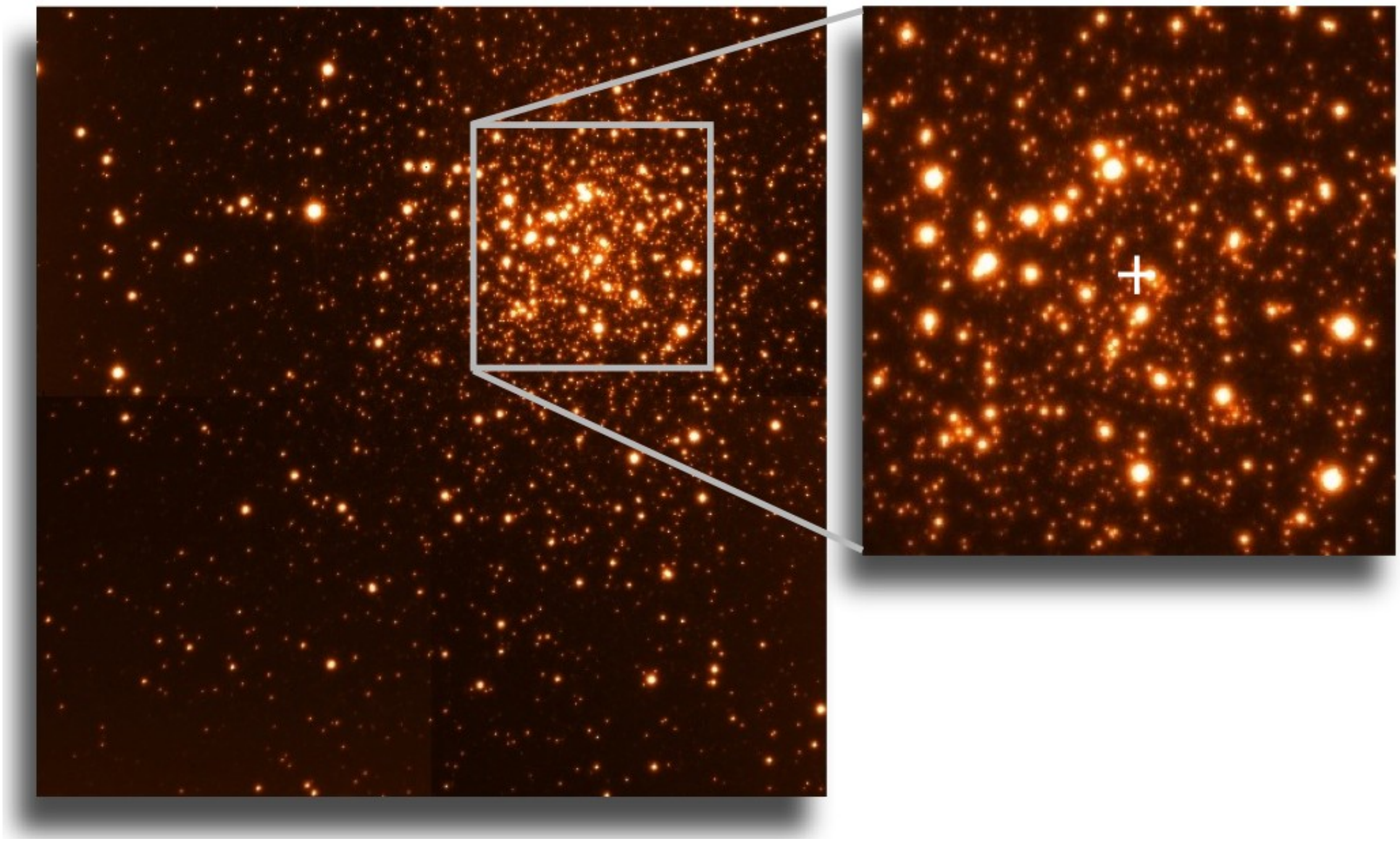

**Fig. 1. MAD image of Terzan 5 in the *K* band.** Observations were performed at the ESO-VLT (Paranal, Chile) on August 2008, through *J* and *K* filters. Exposure times were about two minutes in each filter. Shown is the best image obtained in the *K* band (the image size is 1'x1', north is up, east is left). The measured full-width at half-maximum (FWHM) of stars is 0.1", the Strehl ratio ranges between 15% and 24% over the entire field of view. The quality of the *J* image is slightly worse (FWHM ≈ 0.24" and Strehl ratio below 10%), but still much better than normally obtained with ground-based observations. A small (16" x 16") portion of the *K* image sampling the very central region of Terzan 5 is shown magnified. The cluster centre of gravity (marked with the white cross) has been determined by averaging the positions of the resolved stars and following the same procedure adopted in previous studies[25]. It is located at right ascension $\alpha$ = 17 h 48 m 4.85 s, declination $\delta$= –24° 46' 44.6", which is ~ 3" south-east from the centre listed in the most commonly adopted globular cluster catalogue[14], but in good agreement (within the errors $\Delta\alpha \approx \Delta\delta \approx 0.5$") with the determination obtained from HST-NICMOS observations[9]. The barycenters of the two horizontal branch populations are coincident with the gravity centre within the errors.

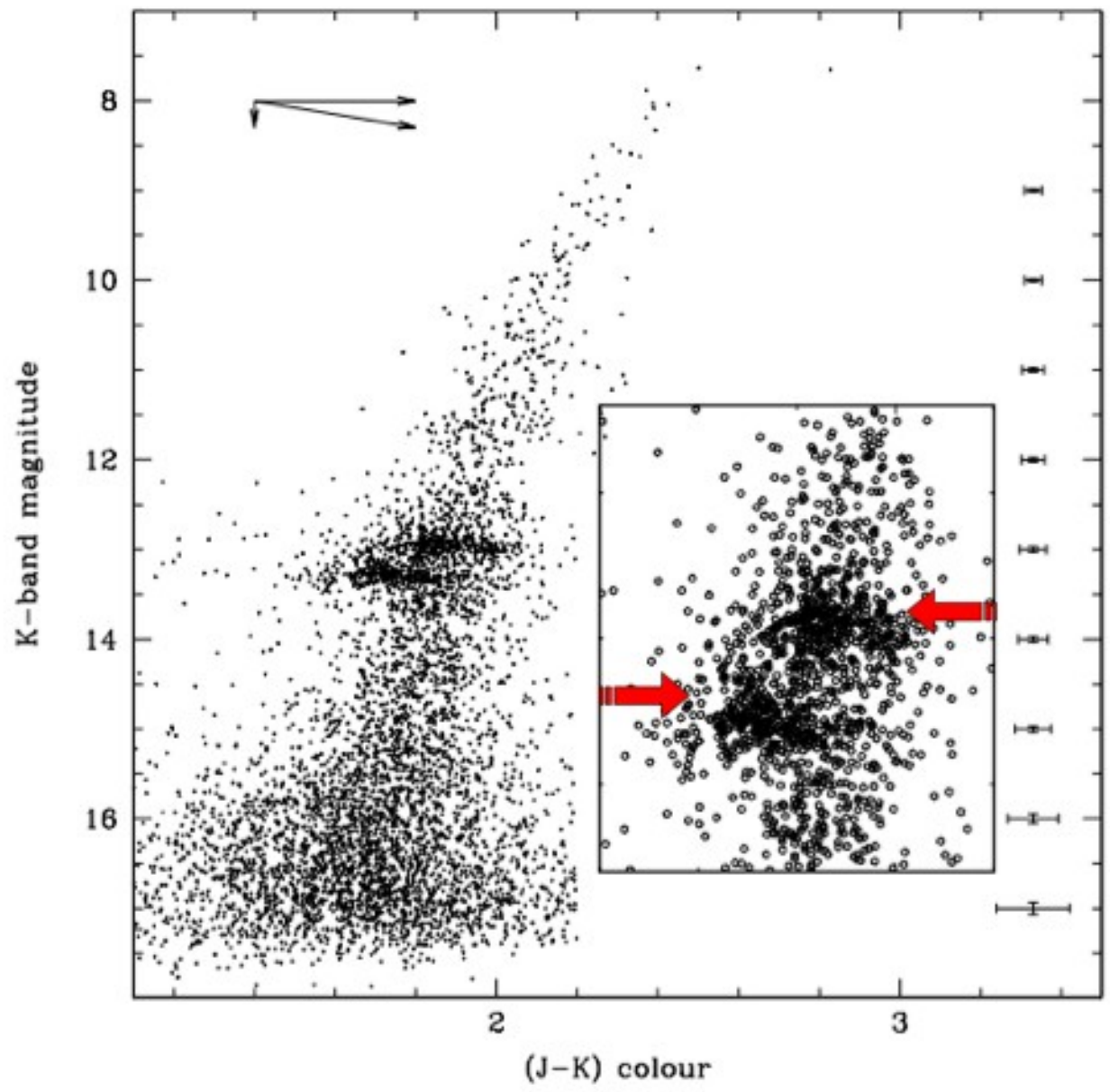

**Fig. 2. The two horizontal branch clumps of Terzan 5.** Main panel, MAD (*K, J*−*K*) CMD of the central region of Terzan 5. Inset, magnified view of the horizontal branch region, with the two horizontal branch clumps marked with red arrows. Terzan 5 is heavily obscured by thick clouds of dust (this effect is commonly called "reddening") intervening between the system and the observer, in a way that strongly depends on the direction of the line of sight ("differential reddening")[7,8]. The effect of reddening on the *K* magnitude and the *J-K* colour is indicated by the reddening vector plotted in the main panel. Several tests have been performed on the images and the catalogue to exclude any possible spurious effect from the instrument or the reduction procedure. Stars in the two clumps do not show any peculiar spatial distribution on the detector. Moreover, the two clumps are not spuriously produced by the variation in size and shape of the Point Spread Function, or the local level of the background. Error bars (1 s.e.m ) are plotted at different magnitude levels.

The contamination from Galactic bulge field stars in this CMD is negligible. In the 1 arcmin$^2$ field of view of MAD, we estimate (Supplementary Information) that 11 and 8 field stars should contaminate the faint and bright horizontal branch selection boxes (while we count 299 *FHB* stars and 310 *BHB* stars in the entire MAD sample).

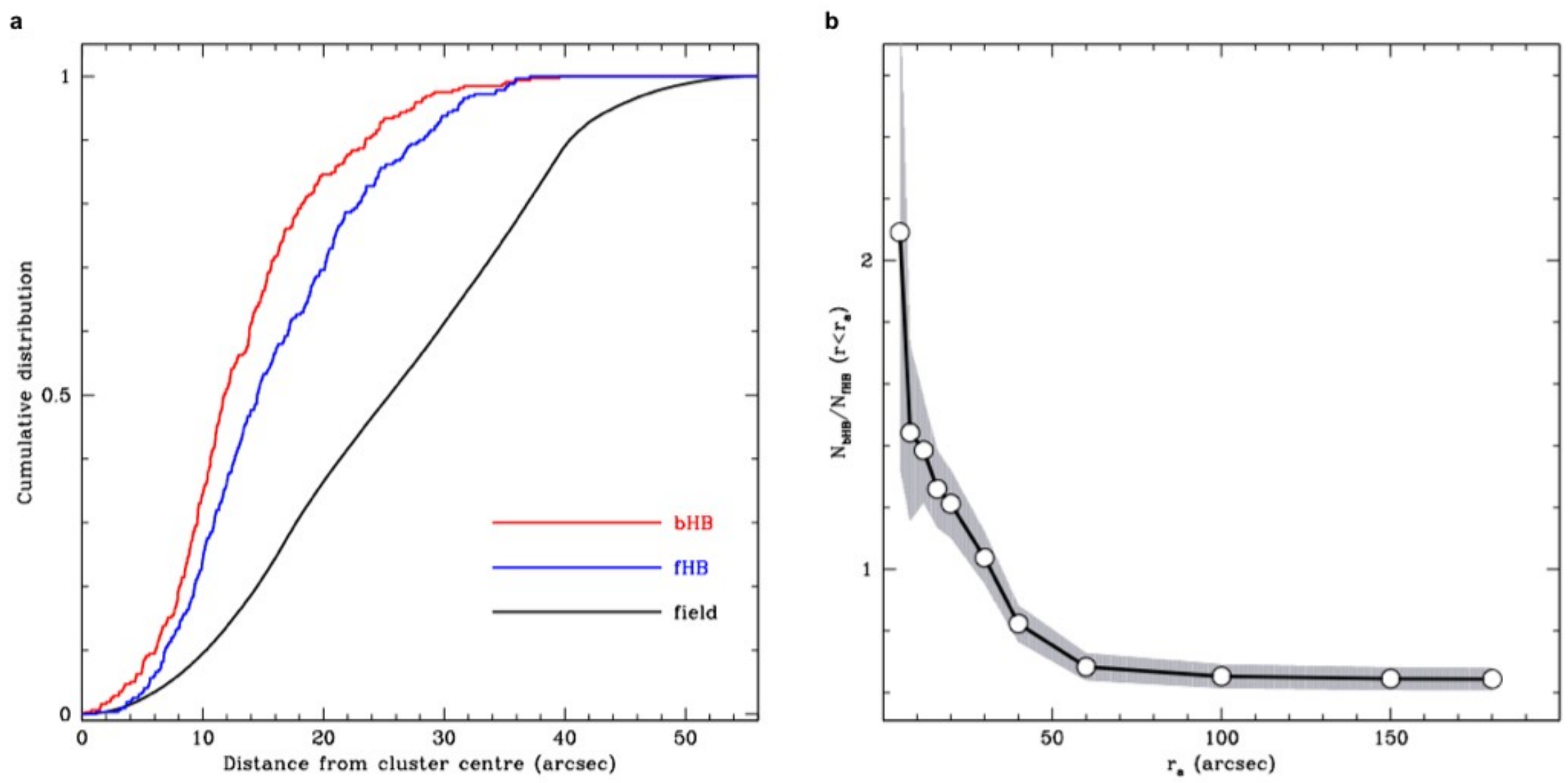

**Fig. 3. Radial distribution of the two horizontal branch populations in Terzan 5.**

**a,** Cumulative radial distribution of the observed *BHB* stars (red line) and the *FHB* population (blue line), compared to that of field stars (solid black line), as a function of the projected distance from the cluster centre of gravity. The field distribution has been obtained from a synthetic sample of 100,000 points uniformly distributed in *X* and *Y* over the MAD field of view.

**b,** Ratio between the number of observed *BHB* and *FHB* stars computed over areas of increasing radius, $r_a$. Points with $r_a$ < 30'' refer to the MAD sample, those corresponding to larger radii have been computed by also using the ACS data. The grey area around the black line represents the 1$\sigma$ uncertainty region. *BHB* stars are substantially more numerous than *FHB* stars in the cluster centre and they rapidly vanish at $r_a$ > 50''.

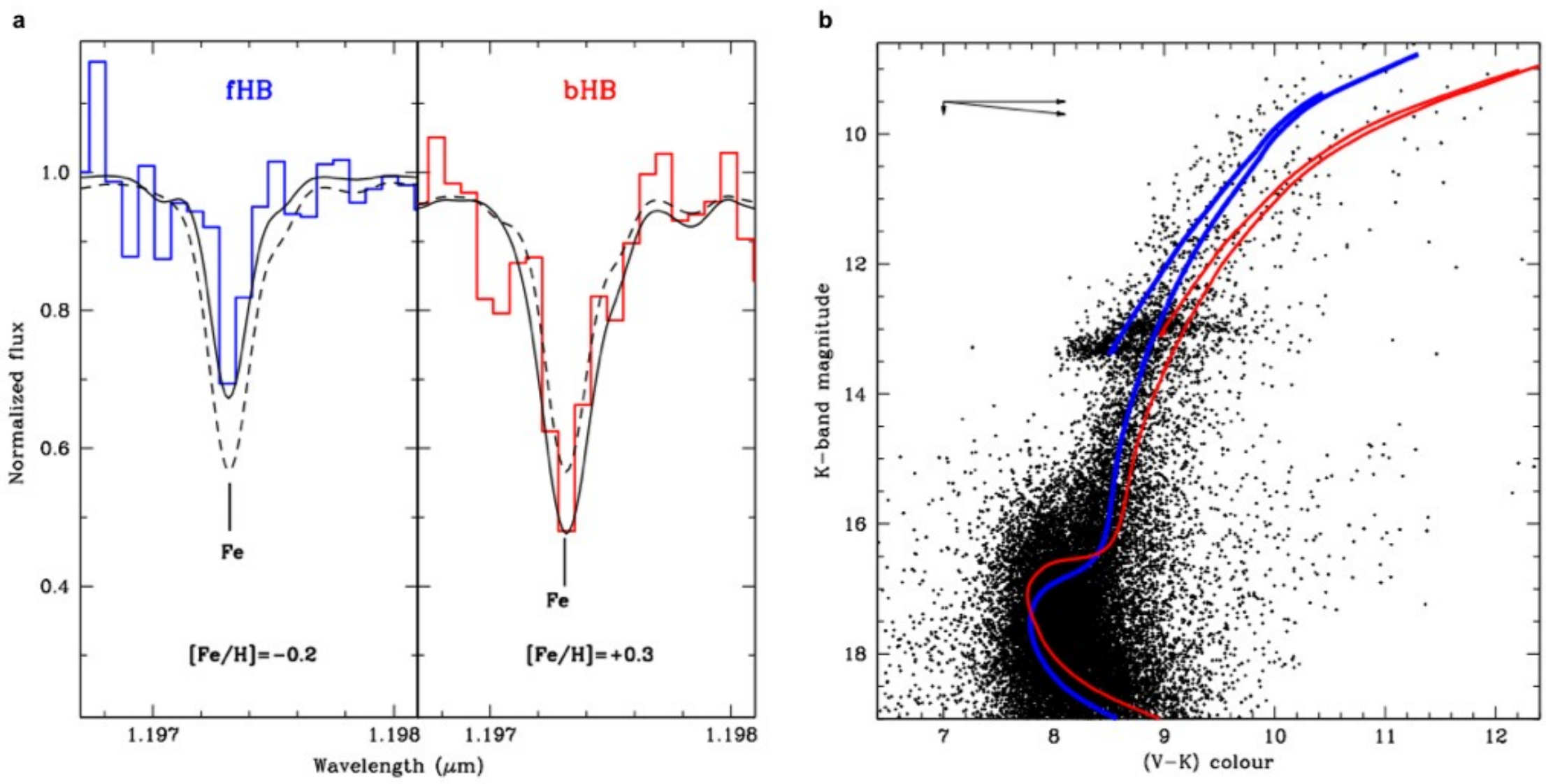

**Fig. 4. Iron abundance and ages of the two populations.**

**a**, Combined *J*-band spectra near the 1.1973 μm iron line for three *FHB* (left) and three *BHB* (right) stars, as obtained with NIRSPEC at *Keck II* on 2 July 2009 (coloured lines). The measured equivalent widths of the lines and suitable spectral synthesis[12] yield iron abundances [Fe/H] ≈ –0.2 ± 0.1 and [Fe/H] ≈ +0.3 ± 0.1, respectively. The black solid lines correspond to the best-fit synthetic spectra obtained for temperatures and gravities derived from evolutionary models reproducing the observed colours of the horizontal branch stars: $T_{eff}$ = 5000 K and log $g$ = 2.5 for the *FHB* stars, $T_{eff}$ = 4500 K and log $g$ = 2.0 for the *BHB* stars. For sake of comparison, we also plot (as black dashed lines) the synthetic spectra obtained by adopting the same atmospheric parameters, but [Fe/H] = +0.3 for the *FHB* and [Fe/H] = –0.2 for the *BHB*.

From the measured spectra we also derived the stellar radial velocities and found an average value of –85 km $s^{-1}$ ($\sigma$ = 9 km $s^{-1}$) and –85 km $s^{-1}$ ($\sigma$ = 10 km $s^{-1}$) for the *FHB* and *BHB* stars, respectively (the typical uncertainty on the individual measure is of the order of 3 km $s^{-1}$). These values are fully consistent with the previously measured radial velocities of four giants ($V_r = -93\pm2$ km s$^{-1}$)[12] and the value ($V_r = -94\pm15$ km s$^{-1}$) listed for Terzan 5 in the currently adopted globular cluster catalogue[14]. This observational fact confirms that the horizontal branch stars for which we have secured spectra are cluster members, and suggests that there is no significant kinematical difference between the two populations.

**b**, ($K$, $V-K$) CMD of Terzan 5 obtained by combining VLT-MAD and HST-ACS data corrected for differential reddening. Two isochrones[26] with [Fe/H] = −0.2 (heavy element mass fraction Z=0.01, and helium mass fraction Y=0.26) and $t$ = 12 Gyr (blue line), and with [Fe/H] = +0.3 (Z=0.03, Y=0.29) and $t$ = 6 Gyr (red line) are overplotted to the data by adopting a colour excess[8] $E(B-V)$ = 2.38±0.05 and a distance[8] $d$ = 5.9±0.5 kpc. Note that the CMD cannot be reproduced by two isochrones with the measured metallicities and the same age. Owing to the large scatter at the turn-off level, we estimate that the uncertainty on the younger component age is about 2 Gyr.

# Supplementary Information

**Field contamination**: In order to estimate the impact of field star contamination on the selected horizontal branch populations, the MAD and ACS samples have been supplemented with optical (*V* and *I*) wide-field (30' × 30') images retrieved from the ESO-Archive. Those horizontal branch stars, as selected in the MAD *(K, J–K)* plane, which are in common with the ACS sample have been used to define the *FHB* and *BHB* selection boxes in the *(I, V– I)* CMD. By counting the number of stars beyond the cluster tidal radius and falling within the horizontal branch selection boxes in the wide-field sample we estimate that 11±3 and 8±3 stars arcmin$^{-2}$ may contaminate the faint and the bright horizontal branch populations, respectively. Since we count 809 *FHB* stars and 521 *BHB* stars in the combined MAD and ACS data set, which samples an area of ~ 11.3 arcmin$^{2}$, the above estimates indicate that ~ 16% of the observed stars are field contamination: this yields ~ 685 true *FHB* stars and 450 *BHB* stars. By considering that the MAD+ACS data set samples 88% of the total cluster luminosity, we estimate that the entire cluster contains a total number of ~800 *FHB* stars and 500 *BHB* stars.

**Proper motions:** To further verify the physical association between the *FHB* and *BHB* populations, we have computed the proper motions of the horizontal branch stars using archived NICMOS *J* images[9] and our own MAD *K* image, which are separated by a temporal baseline of ≈ 10 yrs. The NICMOS field of view (19” x 19”) is smaller than that of MAD. Hence the analysis has been carried out only on a subset of the horizontal branch populations: we counted 175 *FHB* stars and 130 *BHB* stars in the common region. By using the *FHB* as reference to place the two images into the same astrometric system, we computed (a) the residuals between the positions of the *FHB* stars during the two epochs (this sets the internal accuracy of the measures), and (b) the proper motions residuals for the *BHB* population, which traces the relative motion between the two sets of stars. The results of this analysis can be summarized as follows:

1. The systemic tangential velocities of the two sets of stars are identical within the uncertainties. In particular, the difference in tangential velocity is ≈ 0.05 ± 1.08 mas yr$^{-1}$

, which corresponds to 0.0 ±30 km $s^{-1}$, adopting a distance $d = 5.9$ kpc for Terzan 5.

2. The dispersion in tangential velocity is 1.08 mas $yr^{-1}$ for the *FHB* population and 1.25 mas $yr^{-1}$ for the *BHB*. This must be compared with the tangential dispersion of the red clump stars in the Galactic field, which amounts to 3.7 mas $yr^{-1}$ (following the Besancon Galactic model[27] in the direction of Terzan 5). Hence both the *FHB* and the *BHB* populations are significantly dynamically colder than the surrounding Bulge: $\sigma \approx$ 25 km $s^{-1}$ vs. $\sigma \approx 105$ km $s^{-1}$.

3. By deconvolving the tangential dispersion of the *FHB* population (which corresponds to the observational noise) from that of the *BHB* population, we find that the true one-dimensional dispersion of the latter is $\sigma \approx 13$ km $s^{-1}$, in good agreement with the velocity dispersion of the cluster in the radial direction ($\sigma \approx 10$ km $s^{-1}$).

All the above results provide further support to the conclusion that both horizontal branch populations do belong to Terzan 5.

**Implications for the population of millisecond pulsars (MSPs):**

The scenarios discussed in the paper on the possible origin of Terzan 5 have direct implications for the exceptional population of MSPs discovered in this cluster[28]. In fact, if Terzan 5 is the relic of a more massive structure, the deeper potential well of the original system could have favoured the retention of a larger number of (supernova kicked) neutron stars.

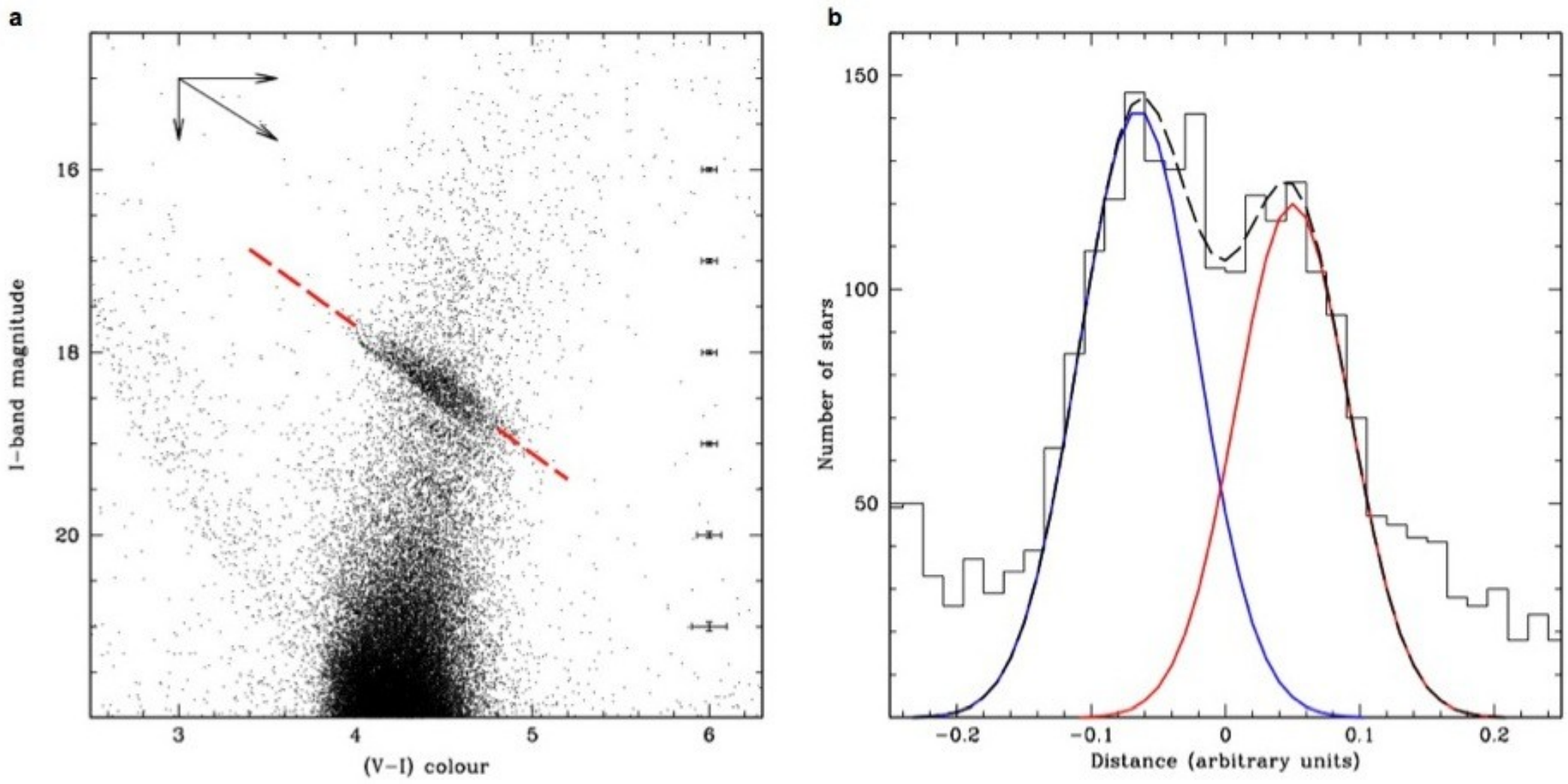

**Supplementary Fig. 1. Optical (*I, V−I*) CMD of Terzan 5 from deep HST images.**

**a**, This CMD demonstrates the difficulty affecting the optical studies of star clusters located in the direction of the Galactic Bulge. In fact, these clusters suffer for heavy dust obscuration, and the reddening effect is to distort all the evolutionary sequences in the CMD: in particular, the two horizontal branch clumps of Terzan 5 appear to be stretched along the direction of the reddening vector (shown in the upper left). Still, two different parallel structures are distinguishable above and below the dashed line, which is parallel to the reddening vector. Error bars, 1 s.e.m are plotted at different magnitude levels.

**b**, The distribution of the horizontal branch star geometrical distances from the dashed line marked in panel **a** is shown. Two well-defined peaks corresponding to the *BHB* and the *FHB* are clearly visible and nicely reproduced by two Gaussians (in red and blue, respectively; the black dashed line is the combination of the two Gaussian distributions).

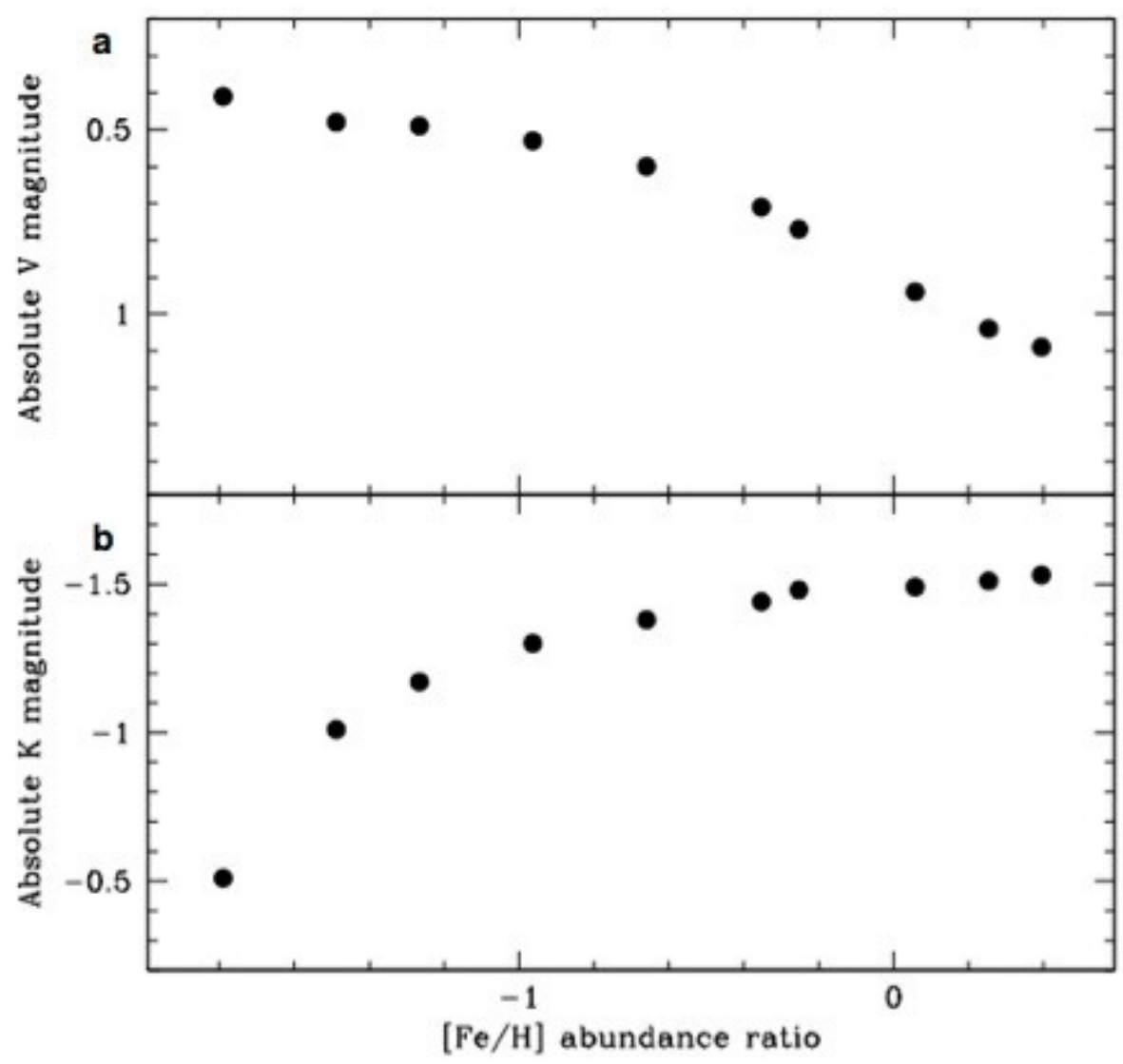

**Supplementary Fig. 2. Theoretical expectations for the dependence of the horizontal branch magnitude level on stellar metallicity.**

**a,** The prediction of theoretical stellar models[26] for the horizontal branch magnitude level in the *V* band is plotted as a function of metallicity.

**b,** As in panel **a**, but for the *K* band.

As apparent and at odds with the behaviour in the optical bands, a metal-rich population is expected to have a brighter horizontal branch than a metal-poor one in the near-infrared. However, for the two observed horizontal branch clumps (with [Fe/H] = –0.2 and [Fe/H] = +0.3) the expected difference is only $\delta K$ ~0.17 mag, i.e. significantly smaller than the observed one ($\delta$K ~ 0.3 mag). In order to reproduce the observations it is necessary to hypothesize that the metal-rich population is younger than the metal-poor one. Note that in principle also a significant enhancement of Helium abundance could produce more luminous horizontal branch stars. However, even at these metallicities, a high helium content would move the horizontal branch stars at the extreme blue tail of the horizontal branch.

# Supplementary References: